\documentclass{vgtc}                          % final (conference style)
\graphicspath{{figures/}{pictures/}{images/}{./}} % where to search for the images

\usepackage{times}                     % we use Times as the main font
\usepackage{amsmath}
\usepackage{amssymb}
\usepackage{epsfig}
\usepackage[english]{babel}

\onlineid{0}

\vgtccategory{Research}

\vgtcinsertpkg

\title{Visualizing Lagrangian Heat Transport Paths and Density Structures in Unsteady Heat Transfer}

\author{Besm Osman\thanks{e-mail: b.osman@tue.nl} %
\and Andrei Jalba\thanks{e-mail: A.C.Jalba@tue.nl} %
\and Michel Speetjens\thanks{e-mail: M.F.M.Speetjens@tue.nl}
\and Anna Vilanova\thanks{e-mail: A.Vilanova@tue.nl}}
\affiliation{\scriptsize Eindhoven University of Technology }

\teaser{
  \centering
  \includegraphics[width=\linewidth]{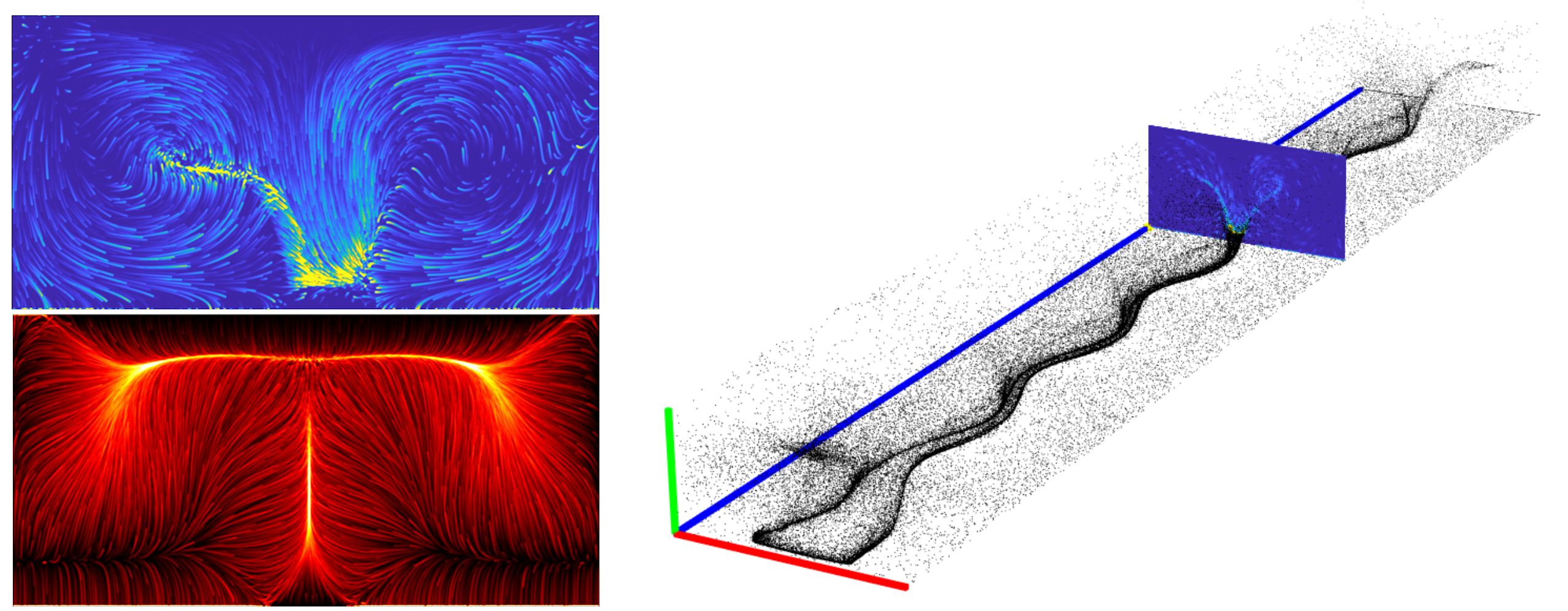}
  \caption{Visualization of thermal transport structures using our proposed method. Left: finite-time density structures with transport paths revealed by particle path accumulation. Right: the 3D spacetime domain along which particles are integrated.}
  \label{fig:teaser}
}

\newcommand{\revision}[1]{#1}

\newcommand{\xvec}[1]{\mbox{\boldmath$#1$}}
\newcommand{\xSpacetime}[0]{\mathbf{x_*}}
\newcommand{\Qspace}{\xvec{Q}'_*}

\abstract{
Convective heat transfer is traditionally visualized from a Eulerian perspective using scalar temperature fields, offering limited insight into the underlying transport mechanisms. A Lagrangian view, analogous to mass transport along fluid paths, can reveal coherent structures and transport routes invisible from a Eulerian view of temperature. However, heat transport is \revision{aperiodic and non-conservative, hampering the application of fluid mixing and transport visualization techniques, developed primarily for time-periodic, conservative transport}. We present a particle-based visualization technique that addresses these challenges by advecting massless particles along a time-reparameterized spacetime formulation of thermal transport, accumulating path contributions to reveal coherent transport routes and finite-time attracting and repelling structures that conventional methods cannot show.
} % end of abstract

\keywords{Scientific visualization, Flow visualization, Lagrangian methods, Coherent structures.}

\begin{document}

%% The ``\maketitle'' command must be the first command after the
%% ``\begin{document}'' command. It prepares and prints the title block.

%% the only exception to this rule is the \firstsection command
\firstsection{Introduction}
\maketitle

Heat transfer in fluid flows (i.e., convective heat transfer) is relevant for a wide range of practical applications including the conventional processing industry, pharmaceutical devices, thermal-storage units for sustainable energy systems, and high-precision equipment. Convective heat transfer is traditionally visualized as scalar maps of temperature fields and investigated with heat transfer coefficients characterizing the heat transfer rate at fluid-solid interfaces~\cite{incroperaFundamentalsHeatMass2007, bejanConvectionHeatTransfer2013}. However, this view offers only limited insight into the actual physical transport mechanisms: temperature fields reveal only the thermal distribution at a particular instant, not where heat is moving to or where it originated, and heat-transfer coefficients provide no insight into thermal transport pathways. Meaningful visualization of convective heat transfer would instead reveal the thermal paths themselves, the coherency of transport along them, and the structures that emerge from this transport.

Speetjens proposed an alternative Lagrangian approach to convective heat transfer~\cite{speetjensGeneralisedLagrangianFormalism2012} that relies on the notion that convective heat transfer fundamentally is the transport of thermal energy by a net heat flux encompassing the combined effect of fluid motion and molecular diffusion. Convective heat transfer thus becomes the transport of thermal energy along certain paths (``thermal paths''), in a similar way as fluid motion is the transport of fluid parcels by the flow along fluid paths. This mass-transfer analogy admits visualization of convective heat transfer by the topology of such thermal paths using well-established Lagrangian methods and concepts from flow visualization and analysis of mixing by the topology of fluid paths~\cite{speetjensLagrangianTransportChaotic2021}. However, two essential differences between fluid and thermal transport render the utilization of existing Lagrangian methods for thermal transport non-trivial. Namely, heat transport is (i) non-conservative \revision{(i.e., net heat flux is divergent)}, and (ii) aperiodic, whereas fluid transport is conservative and generally periodic \revision{in time}. Poincaré maps require periodicity that thermal transport does not have. \revision{Lagrangian coherent structure techniques~\cite{hallerLagrangianCoherentStructures2015} can capture transport structures in aperiodic fluid transport but do not reveal the path along which the transport occurs. Their physical interpretation in the context of thermal transport remains an open question.} Existing flow visualization techniques such as Line Integral Convolution and the density-based method of Park et al.~\cite{parkStructureaccentuatingDenseFlow2006} \revision{can reveal flow behavior along with attracting and repelling structures, but only} in steady flows, \revision{resulting in a view of the instantaneous structures in a time-frozen unsteady flow rather than the time-evolving transport structures.}

In this work, we take a first step \revision{toward} capturing coherent structures in convective heat transport by presenting a visualization technique for aperiodic divergent transport that simultaneously captures finite-time density structures and coherency in transport paths. The method works by advecting massless particles \revision{within} a time segment of the spacetime formulation of thermal transport following from the Lagrangian transport formalism. Contributions of particles are then accumulated via a Gaussian kernel \revision{along} path segments \revision{on an accumulation grid}, producing \revision{both still images} and animations revealing coherency of thermal transport routes and density evolution.

\section{Background: Lagrangian thermal formalism}\label{sec-background}
Our work uses the Lagrangian thermal formalism (LTF) proposed by Speetjens~\cite{speetjensGeneralisedLagrangianFormalism2012} to extract the Lagrangian view of the convective contribution of heat transport from standard computational fluid dynamics (CFD) derived temperature evolutions. In this section, we briefly introduce the formalism. The inputs are the flow field $\xvec{u}(\xvec{x},t)$ and temperature field $T(\xvec{x},t)$, where $\xvec{x} \in \mathcal{D}$ denotes position in spatial domain $\mathcal{D}$ and $t$ denotes physical time. All subsequent fields are functions of $(\xvec{x},t)$; arguments are dropped for brevity. The first step of the LTF involves the decomposition of the temperature field as
\begin{align}
T(\xvec{x},t) = \widetilde{T}(\xvec{x},t) + T'(\xvec{x},t),
\label{Temperature1}
\end{align}
with $\widetilde{T}$ the temperature evolution in the case of a stagnant fluid, i.e., only due to conduction, $\xvec{u}=\xvec{0}$. The stagnant temperature evolution is readily available via CFD simulations. Contribution $T'$, by definition, incorporates the impact of the flow $\xvec{u}$, i.e., $T'\neq0$ only if $\xvec{u}\neq\xvec{0}$. This contribution is governed by the transport equation
\begin{align}
\frac{\partial T'}{\partial t} +
\xvec{\nabla}\cdot\xvec{Q}' = F,
%\quad
%
\label{Convective1}
\end{align}
revealing that $T'$ emanates from the interplay of two fundamental transport mechanisms: net heat flux $\xvec{Q}' = \xvec{u}T' - \alpha\xvec{\nabla}T'$ and source $F = -\xvec{u} \cdot \xvec{\nabla}\widetilde{T}$. \revision{The net heat flux $\xvec{Q}'$, in turn, represents the combined heat transport due to the interplay of convection ($\xvec{u}T'$) and conduction ($- \alpha\xvec{\nabla}T'$) relative to the reference temperature $\widetilde{T}$.}

Transport equation~\eqref{Convective1}~\cite{incroperaFundamentalsHeatMass2007} is the backbone of the visualization and Lagrangian analysis of convective heat transfer enabling its description as the motion of a fluid. Substitution
\begin{align}
T'= \rho,\quad \xvec{Q}'=\xvec{M},\quad F = F_\rho,
\label{MassFlux1}
\end{align}
with $\rho$ the fluid density, $\xvec{M}$ the mass flux and $F_\rho$ a volumetric mass source/sink (e.g., due to chemical reactions), translates \eqref{Convective1} into the conservation law for mass \cite{incroperaFundamentalsHeatMass2007}.
The discussion below is in terms of this analogy; ``fluid/flow'' properties $(\rho,\xvec{M},F_\rho)$ hereby represent their thermal counterparts $(T',\xvec{Q}',F)$ following \eqref{MassFlux1}.

Convective heat transfer in the current LTF corresponds with the motion of fluid particles, density $\rho$, at velocity $\xvec{v} = \xvec{M}/\rho$ along trajectories $\xvec{x}(t)$ described by the 
kinematic equation and associated formal solution
\begin{align}
\frac{d\xvec{x}}{dt} = \xvec{v}\left(\xvec{x}(t),t\right) \quad\Rightarrow\quad \xvec{x}(t) = \xvec{\Phi}_t(\xvec{x}_0),\nonumber\\
\xvec{\Phi}_t(\xvec{x}_0)\equiv \xvec{x}_0 + \int_0^t\xvec{v}\left(\xvec{x}(\eta),\eta\right)d\eta,
\label{KinEq1}
\end{align}
where $\xvec{x}_0 = \xvec{x}(0)$ denotes the initial position. The thermal equivalent of $\xvec{v}$ corresponds to $\xvec{Q}'/T'$ which, due to $T'$ having both positive and negative regions, is numerically unfavorable to integrate where $T'$ approaches zero. Kinematic equation~\eqref{KinEq1} has an equivalent formulation

\begin{align}
\frac{d\xvec{x}}{d\xi} = \xvec{M}\left(\xvec{x}(\xi),\revision{t(}\xi\revision{)}\right) \quad\Rightarrow\quad \xvec{x}(\xi) = \xvec{\Phi}_\xi(\xvec{x}_0),
\label{KinEq2}
\end{align}
with fictitious ($\xi$) and real ($t$) time relating via
\begin{align}
t = f(\xi) = \int_0^\xi \rho\left(\xvec{x}(\eta),\eta\right)d\eta.
\label{TimeAxis1}
\end{align}
This equivalence enables determination of the trajectories described by \eqref{KinEq1} via \eqref{KinEq2}, without numerically dividing by $T'$. Important to note is that \eqref{Convective1} and \eqref{MassFlux1} yield
\begin{align}
\frac{d\rho}{dt}\equiv\frac{\partial \rho}{\partial t} + \xvec{v}\cdot\xvec{\nabla} \rho = F_\rho - \rho\xvec{\nabla}\cdot\xvec{v},
\label{NonConservative}
\end{align}
meaning that the generic $F_\rho\neq 0$ implies non-constant particle density (i.e., $d\rho/dt\neq 0$) and, inherently, non-divergence-free flow (i.e., $\xvec{\nabla}\cdot\xvec{v}\neq 0$). This renders system \eqref{KinEq1} non-conservative.

\section{Method}    
\subsection{Overview}
\begin{figure}
    \centering
    \includegraphics[width=.95\linewidth, alt={Illustration showing particles starting in a seeding window moving in spacetime towards an accumulation frame.}]{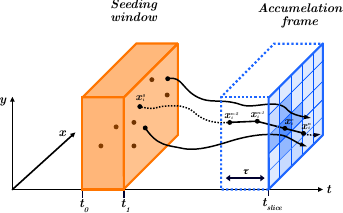}    \caption{Illustration of particle integration in the spacetime domain and accumulation onto a regular grid.}
\label{fig:placeholder}
\end{figure}

The structures emerging from the LTF remain largely unexplored due to the lack of a suitable visualization mechanism. We propose a method that captures these transport structures and paths in two-dimensional unsteady thermal flows \revision{governed by kinematic equation \eqref{KinEq2}}. Our method advects massless particles along the spacetime representation of the thermal flow following from the LTF (illustrated in Figure~\ref{fig:placeholder}). The inputs are identical to those of the LTF: A time-dependent temperature field with flow $T$ and without flow $\widetilde{T}$, from which $T', \xvec{Q}'$ and $\xvec{F}$ are derived (see equations~\eqref{Temperature1} and~\eqref{Convective1}). Particles are seeded within a temporal window of the domain (see Section~\ref{sec-seeding}), yielding attracting structures (\revision{heat sinks}) under forward time integration and repelling structures (\revision{heat sources}) under backward integration. The particle paths are visualized by accumulating onto a fixed grid via a Gaussian kernel applied to particle path segments (see Section~\ref{sec-accumulation}). This allows finite-time density structures and coherency of transport paths to be visualized either as an image or animation (see Section~\ref{sec-rendering}).

\subsection{Spacetime Integration}\label{sec-integrating}
While $\xvec{Q'}$ is unsteady in the 2D physical domain, \revision{lifting $t$ as a third phase-space coordinate yields an autonomous system $\Qspace = (Q'_x,\, Q'_y,\, T')$ in the spacetime domain}. Massless particles are integrated along this steady thermal flow via fictitious time $\xi$, i.e., $\frac{d}{d\xi}(x,y,t) = \Qspace$, enabling visualization of transport along temporal slices of the spacetime domain. The reparametrization produces trajectories equivalent to those of the thermal flow while avoiding integration issues near $T' = 0$, as shown in Section~\ref{sec-background}. We integrate numerically using fourth-order Runge-Kutta on a trilinear interpolation of $\Qspace$. Each particle $i$ is tracked by its spacetime position $\xSpacetime^n_i = (\xvec{x}^n_i, t^n_i)$, where $\xvec{x}^n_i \in \mathcal{D}$ is the spatial position and $t^n_i$ is the physical time at simulation step $n$. \revision{Forward ($\Delta\xi > 0$) and backward ($\Delta\xi < 0$) integration reveal attracting and repelling structures respectively. Note that the direction of $\xi$ does not necessarily coincide with that of physical time $t$, since $dt/d\xi = T'$, so regions where $T' < 0$ cause forward $\xi$-integration to decrease $t$. This signifies a negative heat flux backward in time, which is equivalent to a positive heat flux forward in time.} Particles move at varying speeds \revision{through $t$}, affecting density evolution. This cannot be resolved by adaptive time steps of $\Delta\xi$ without reintroducing numerical issues near $T' = 0$; instead, we address the density bias through our seeding strategy.

\subsection{Seeding}\label{sec-seeding}
Seeding of particles has three goals: well-distributed particles for path visualization, physically meaningful density evolution, and visual insight into transport structures. These goals conflict. Seeding proportional to $F$ captures physical density evolution but leaves regions of low $|F|$ sparse, obscuring transport paths. Uniform seeding distributes particles well initially, but introduces domain-shape bias over time as particles migrate toward transport routes, leaving much of the domain empty and causing artificial sharp transitions to appear~\cite{parkStructureaccentuatingDenseFlow2006}.

We address these conflicts as follows. Particles are seeded uniformly over the spacetime domain using a Halton sequence for quasi-random seed generation, where each particle $i$ is assigned weight $w_i$ initialized to $F$ at its initial position and time, ensuring well-distributed particles. \revision{The weighting provides an initial importance measure used in the subsequent accumulation to improve the visibility of transport structures.} To maintain \revision{visibility of} transport structures over longer integration times, particles are reseeded at a rate proportional to elapsed fictitious time. The constant particle influx and uniform removal stochastically stabilize the particle distribution, preventing particles from collapsing towards stagnation regions. While reseeding sacrifices physical accuracy of the density evolution, it reveals a wider range of transport structures that would otherwise become invisible at long integration times. We specify seeding to occur over a finite time seeding window $[t_0, t_1]$ (see Figure~\ref{fig:placeholder}) to allow the capture of finite-time structures.

\subsection{Path Segment Accumulation}\label{sec-accumulation}
Particle contributions within time $\tau$ are accumulated on a two-dimensional regular grid $A(\xvec{p})$, where $\xvec{p} \in \mathcal{D}$, using a 2D spatial Gaussian kernel. Similar to Kernel Density Estimation~\cite{chenTutorialKernelDensity2017} and Smoothed Particle Hydrodynamics~\cite{monaghanSmoothedParticleHydrodynamics2005} techniques, a Gaussian kernel allows for the estimation of a density field from limited samples. The kernel is isotropic with width $\sigma$, providing a tradeoff between smoothness and detail of the density field.

Accumulating discrete particle positions within $\tau$ results in gaps when the particle trajectory segments defined by $\Delta\xi\cdot|\xvec{Q}'|$ exceed the size of a sample grid cell. Other work in steady flow visualization resolves this by normalizing the vector field \cite{parkStructureaccentuatingDenseFlow2006, han-weishenNewLineIntegral1998, vanwijkImageBasedFlow2002}. However, this is unsuitable here, as the varying divergence of the transport fields is physically meaningful. Instead, we accumulate segments $[\xSpacetime^{n-1}_i, \xSpacetime^{n}_i]$ without modification of the transport field rather than discrete positions. The contribution of a segment to a sample position $\xvec{p}$ of the regular grid is defined by:
\begin{equation}
a(\xvec{p}, \xvec{x}^{n-1}_i, \xvec{x}^n_i) = f(\mu_i) \cdot 
\left(1 - \frac{|t_i^c - t_\text{slice}|}{\tau}\right)\cdot e^{-\frac{||\xvec{x}^c_i - \xvec{p}||^2}{2\sigma^2}}
\end{equation}
where $\xvec{x}^c_i$ is the closest spatial point along the segment to $\xvec{p}$ with the corresponding physical time $t_i^c$, see Figure~\ref{fig:placeholder}. The formula consists of three terms: the contribution \revision{at} $\xvec{p}$ from the Gaussian kernel centered at the closest segment point $\xvec{x}^c_i$ ; a time factor that reduces contribution for particle positions proportional to the difference of the physical time $t_i^c$ from the visualized time slice $t_\text{slice}$; and a fade-in factor $f(\mu_i)$, a smoothly increasing function from 0 to 1 over the particle lifetime $\mu_i$ (e.g., smoothstep), preventing abrupt contributions from newly reseeded particles.

Accumulating segments directly produces overlap artifacts, since the accumulation effectively adds tube segments with varying opacity. The beginning of each segment overlaps with the end of the previous one, resulting in double accumulation at segment boundaries. This can be resolved by subtracting the contribution of the previous segment $[\xSpacetime^{n-2}_i, \xSpacetime^{n-1}_i]$ from that of the current segment. However, full subtraction eliminates contributions from slow-moving particles. While this is beneficial for path visualization, it would reduce visibility of density structures. We therefore subtract proportionally to $l = 2\sigma^{-1}|\xvec{x}^{n-1}_i - \xvec{x}^n_i|$, reducing the subtraction for slow-moving particles based on segment length relative to the kernel width $\sigma$. The total accumulation $A(\xvec{p})$ sums contributions of all particle path segments within $\tau$, weighted by $w_i$ as defined in Section~\ref{sec-seeding}. The accumulation field $A(\xvec{p})$ is defined as
\begin{equation}
A(\xvec{p}) = \sum^{m}_{i=1} w_i \cdot \max \left( a(\xvec{p}, \xvec{x}^{n-1}_i, \xvec{x}^{n}_i) 
- l \cdot a(\xvec{p}, \xvec{x}^{n-2}_i, \xvec{x}^{n-1}_i), 0\right)
\end{equation}
where $m$ is the total number of particles. Since the spacetime formulation following from the LTF is \revision{autonomous}, the accumulation field at a fixed temporal window defined by $t_\text{slice}$ and $\tau$ captures the same physical dynamics regardless of fictitious integration time $\xi$. Blending between successive frames therefore accumulates contributions from particles at different stages of their fictitious integration, all depicting the same physical time slice. This is achieved by applying an exponential decay rate $d$ at each frame,
\begin{equation}
A(\xvec{p}) \gets A(\xvec{p})/(1+d),
\end{equation}
so that older segment contributions fade over time, revealing transport paths. When rendered as an animation, the fading path reveals directionality without altering the underlying physical dynamics.

\subsection{Rendering}\label{sec-rendering}
The accumulation field $A(\xvec{p})$ is scaled by a global factor and rendered to a texture by applying a colormap as a transfer function. For performance, segment contributions are computed only over samples within the spatial bounding box of each segment, avoiding iteration over the full accumulation grid. The accumulation field is rendered to a texture at a resolution that is an integer multiple of the \revision{accumulation} grid, allowing bilinear interpolation to smooth the output without additional computation. \revision{The full visualization method is implemented in FlowExplainer~\cite{flowexplainer}.}

\section{Results}
\begin{figure*}
    \centering
    \includegraphics[width=.82\linewidth, alt={Four-panel figure. Top left: temperature with flow shows a rising plume in the center. Bottom left: temperature without flow shows a horizontally uniform gradient. Top right: attracting structures showing convective transport paths. Bottom right: repelling structures showing where particles accumulate, colored by particle path density.}]{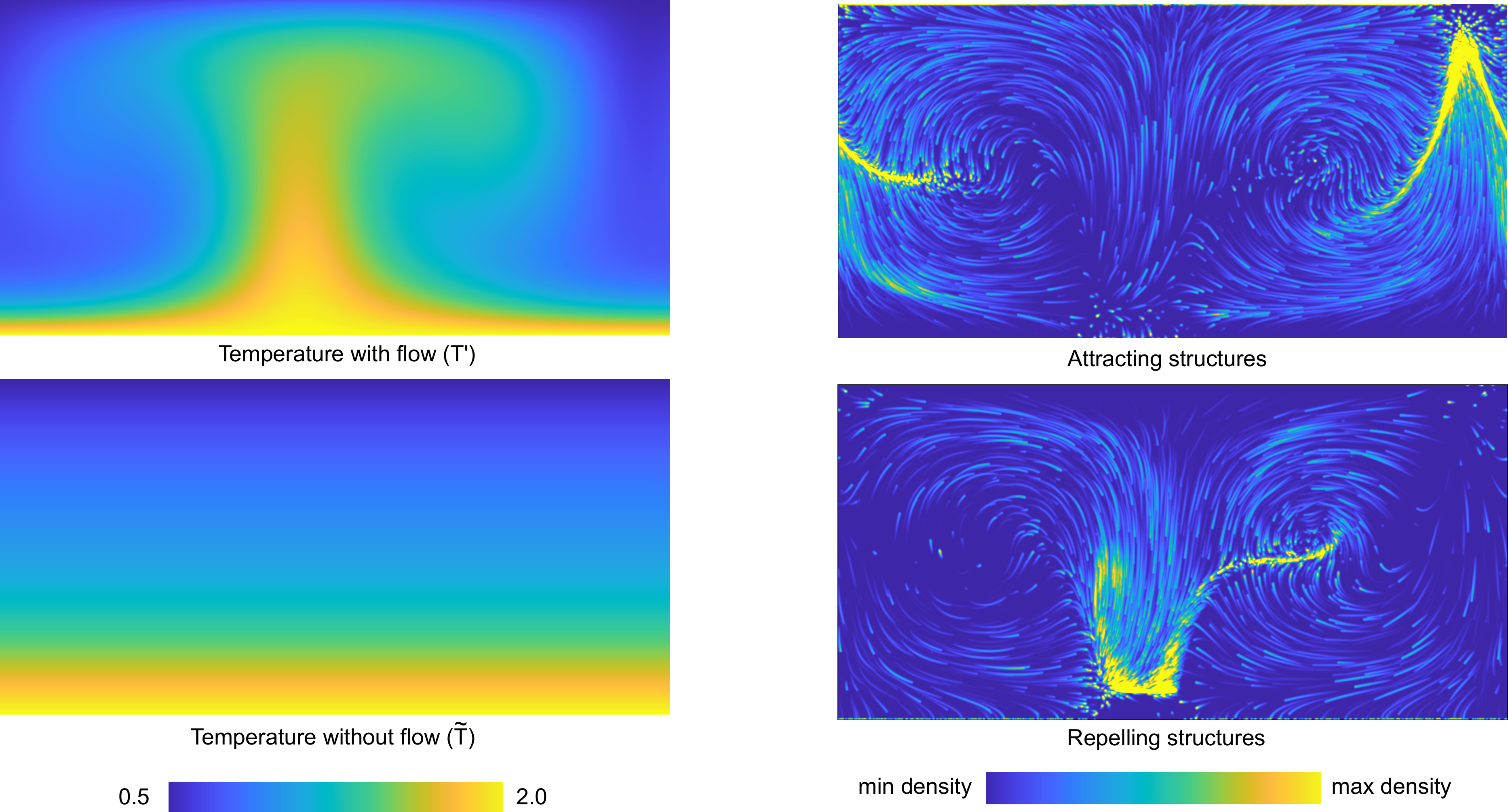}
    \caption{Temperature field and Lagrangian visualization of  \revision{net} convective \revision{heat} transport structures. Left: Eulerian view of the temperature distribution with flow $T'$ (top) and without flow $\widetilde{T}$ (bottom). Right: attracting (top) and repelling (bottom) structures along with \revision{net convective heat} transport paths, \revision{colored by} particle path density. All panels use the \textit{parula} colormap~\cite{matlabParula}} 
    \label{fig:temp}
\end{figure*}

\subsection{Dataset}
We consider a 2D time-dependent convective heat transfer case introduced in~\cite{speetjensGeneralisedLagrangianFormalism2012}. The domain is $\mathcal{D} = [0,1] \times [0,\frac{1}{2}]$ with periodic boundary conditions \revision{in the $x$-axis}. Heat transfer \revision{is} driven by a temperature difference between a ``hot'' bottom wall and a ``cool'' top wall, \revision{with} a time-periodic solenoidal velocity field \revision{consisting} of two adjacent counter-rotating vortices undergoing horizontal oscillation, with $\mathrm{Pe}=100$. \revision{Consistent with~\cite{speetjensGeneralisedLagrangianFormalism2012}, this value balances convection and diffusion in $\mathbf{Q}'$ without either dominating.} Temperature fields are computed using a custom spectral CFD solver. Full details are given in \cite{speetjensGeneralisedLagrangianFormalism2012, speetjensVISUALIZATIONHEATTRANSFER2011}.

\subsection{Finite-Time Density Structures}\label{sec-density-structures}
Figure~\ref{fig:temp} left shows the conventional Eulerian view of temperature evolution at a particular time slice, along with the stagnant case $\widetilde{T}$ without convection. The right \revision{column} shows the result of applying our visualization technique with $t_0=0$, $t_1=t_\text{slice}$, $\tau = 0.05$, \revision{$\sigma=0.04$, $d=0.05$ and $16{,}000$ particles.} The forward integration shows attracting structures where particles concentrate (i.e., in yellow), indicating the regions of convective heating and the transport directions along which heat arrives there. The backward integration reveals the corresponding repelling structures, identifying regions of convective cooling. These structures are not visible in the temperature fields in Figure~\ref{fig:temp} \revision{nor in the fluid velocity field due to the non-trivial relationship between $\xvec{Q}'$ and $\xvec{u}$ (see Section~\ref{sec-background}).}

\subsection{Coherent Structures in Unsteady Transport}\label{sec-unsteady-paths}
\begin{figure}[b!]
\centering
        \includegraphics[width=.90\linewidth, alt={Two images showing coherent structures shaped like the letter T. The top image, from a steady flow, has an unbroken has more pronounced structure. The bottom image, from a periodic flow, has less coherency as transport paths cross.}]{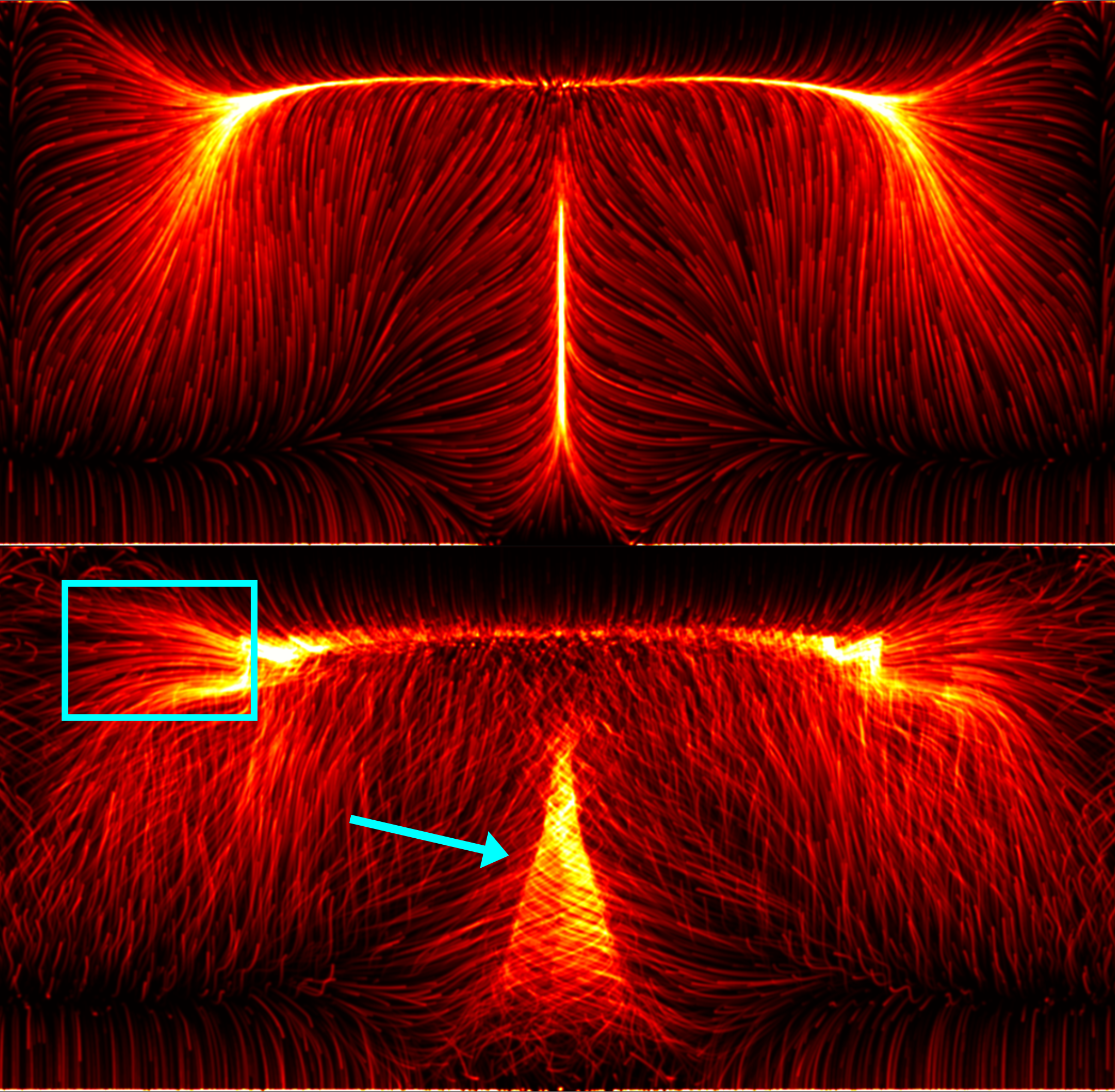}
   \caption{Repelling structures in the diffusion component of heat transport (\textit{hot} colormap~\cite{matlabHot}) driven by a steady (top) and a periodic (bottom) flow. \revision{Although transport is aperiodic in both, coherent paths and thermal structures emerge.}}
    \label{fig:unsteady-paths}
\end{figure}
Beyond visualizing the full heat flux $\xvec{Q}'$, the method can be applied to individual flux contributions to gain targeted insight. Figure \ref{fig:unsteady-paths} shows the visualization applied to the diffusion contribution $\alpha\xvec{\nabla}T'$ of two flows with differing degrees of unsteadiness, with $\tau$ increased to reveal coherency across a wider temporal slice.

The top case is driven by a steady flow, yet the resulting heat transport is unsteady. The largely non-crossing paths indicate coherent transport structure despite this unsteadiness. The bottom case is driven by an unsteady periodic flow, producing more complex aperiodic transport behavior. Despite the heat transport being aperiodic, the visualization reveals what appear to be periodic structures forming. At the blue arrow, transport rotates around an attracting region while moving upward, suggesting a periodic structure, most clearly visible in the provided supplementary video. Regions of consistent directionality are also visible, as seen in the blue rectangle, where uniform transport directions indicate that flow toward attracting structures varies little over the given timeframe.

\section{Discussion \& Conclusion}
We presented a particle-based visualization technique that reveals finite-time density structures and coherent transport routes in unsteady divergent heat transport, taking a first step toward making coherent structures in unsteady thermal transport visible. Current limitations include the restriction to 2D flows and that large temporal windows $\tau$ can produce clutter in highly unsteady flows. The method \revision{conceptually} generalizes beyond the LTF to other divergent transport problems. Future work includes \revision{gaining deeper insight into} the physical meaning of these structures, extending the method to three-dimensional heat transfer, and applying it to other transport datasets. Overall, the proposed method provides visual insight into convective heat transfer mechanisms that conventional methods cannot reveal.

%% \section{Introduction} %for journal use above \firstsection{..} instead
% This template is for papers of VGTC-sponsored conferences which are \emph{\textbf{not}} published in a special issue of TVCG.

%% if specified like this the section will be committed in review mode
% \acknowledgments{}

%\bibliographystyle{abbrv}

\bibliographystyle{abbrv-doi}
\clearpage
\noindent
\bibliography{template}
\end{document}